\documentclass[12pt]{elsarticle}

\usepackage{amsmath}
\usepackage{graphicx}
\usepackage{subfigure}
\usepackage{color}
\usepackage{dcolumn}
\usepackage{bm}
\usepackage{epstopdf}
\usepackage{hyperref}
\biboptions{sort&compress}

\journal{Phys. Lett. A}

\begin{document}
\begin{frontmatter}

\title{Thermodynamic and critical properties of an antiferromagnetically stacked triangular Ising antiferromagnet in a field}

\author{M. \v{Z}ukovi\v{c}\corref{cor}}
\ead{milan.zukovic@upjs.sk}

\author{M. Borovsk\'{y}}
\author{A. Bob\'ak}

\address{%
Institute of Physics, Faculty of Science, P. J. \v{S}af\'arik University, \\
Park Angelinum 9, 040 01 Ko\v{s}ice, Slovakia
}%

\date{\today}

\cortext[cor]{Corresponding author}

\begin{abstract}
We study a stacked triangular lattice Ising model with both intra- and inter-plane antiferromagnetic interactions in a field, by Monte Carlo simulation. We find only one phase transition from a paramagnetic to a partially disordered phase, which is of second order and 3D XY universality class. At low temperatures we identify two highly degenerate phases: at smaller (larger) fields the system shows long-range ordering in the stacking direction (within planes) but not in the planes (stacking direction). Nevertheless, crossovers to these phases do not have a character of conventional phase transitions but rather linear-chain-like excitations.

\end{abstract}


\end{frontmatter}

\section{Introduction}

A stacked triangular Ising antiferromagnet (STIA) is a geometrically frustrated spin system that has attracted considerable attention over the past several decades~\cite{Berker,Blankschtein,Coppersmith,Heinonen,Kim,Netz1,Netz2,Plumer-1,Plumer0,Plumer1,Plumer2,Plumer3,Bunker,Nagai,Kurata,Koseki,Todoroki,Meloche,Zukovic1,Zukovic2,Liu} due to its frustration-induced intriguing and controversial behavior as well as the fact that it reasonably describes some real magnetic materials, such as the spin-chain compounds $\rm{CsCoX}_3$ (X is Cl or Br) and $\rm{Ca}_3\rm{Co}_2\rm{O}_6$. The model consists of layers of triangular lattices stacked on top of each other thus forming linear chains of spins in the perpendicular direction. The interaction between spins within the chains (or between layers) can be considered to be either ferromagnetic (FSTIA model) or antiferromagnetic (ASTIA model). 

In the absence of an external magnetic field the physics of both systems is the same and, therefore, most of the previous studies chose the FSTIA model for their investigations~\cite{Berker,Blankschtein,Coppersmith,Heinonen,Kim,Netz1,Netz2,Plumer1,Plumer2,Plumer3,Bunker,Nagai,Kurata,Meloche,Zukovic1,Zukovic2,Liu}. In zero field, the system has been found to undergo a second-order phase transition from the paramagnetic (P) to a partially disordered (PD) phase $(M,-M,0)$, with two sublattices ordered antiferromagnetically and the third one disordered. There is a wide consensus that the transition belongs to the 3D XY universality class~\cite{Berker,Blankschtein,Plumer1,Bunker,Meloche} albeit the tricritical behavior has also been suggested~\cite{Heinonen}. Another phase transition at lower temperatures to a ferrimagnetic (FR) phase $(M,-M/2,-M/2)$, with one sublattice fully ordered and two partially disordered has been proposed~\cite{Blankschtein,Netz1,Todoroki} but questioned by several other studies~\cite{Coppersmith,Heinonen,Zukovic2,Borovsky}, which argued that the low-temperature phase is a 3D analog of the 2D Wannier phase.

In the presence of the magnetic field, most of theoretical studies focused on elucidation of peculiar phenomena in magnetization processes observed in the experimental realizations $\rm{CsCoX}_3$ and $\rm{Ca}_3\rm{Co}_2\rm{O}_6$~\cite{Zukovic1,Kudasov1,Kudasov2,Kudasov3,Yao1,Yao2,Yao3,Qin,Soto,Kudasov}. Also critical properties of the FSTIA model have attracted a lot of interest due to phase transitions belonging to a variety of universality classes and multicritical behavior. In particular, the Monte-Carlo Mean-Field theory predicted the phase diagram in the temperature-field plane, with a small region of the PD phase stabilized at higher temperatures and small fields and the remaining part occupied by the FR phase~\cite{Netz1}.The character of the P-PD transition line is concluded as second-order belonging to the XY universality class, however, at higher fields the P-FR transition line is identified as first-order due to its thee-state Potts universality class. The FR-PD is reasoned to belong to the Ising universality class with possible crossover to the first-order behavior at low temperatures and very small fields. Later Monte Carlo simulations confirmed the first-order nature of the P-FR transition, however, suggested that the PD phase is probably destabilized by any finite field and phase transitions at smaller fields were determined to belong to the tricritical universality class~\cite{Plumer3}.       

There have been attempts to also determine the phase diagram of the ASTIA model, which in the presence of the field is expected to differ from the FSTIA model, by the Monte-Carlo Mean-Field~\cite{Netz2} and the Landau~\cite{Plumer-1,Plumer0} theories. Both approaches predicted, besides the high-temperature P-PD line of second-order  transitions, also one~\cite{Netz2} and up to two~\cite{Plumer-1,Plumer0} phase transitions to ferrimagnetic states at lower temperatures which can be first- or second-order of the Ising universality. The goal of the present study is to confront these early results obtained by the above approximate approaches with Monte Carlo (MC) simulations and a finite-size scaling analysis.
 
\section{Model and methods}
\label{model}

\subsection{Model}
\begin{figure}[t!]
\centering
\vspace*{-15mm}
\includegraphics[width=0.5\textwidth]{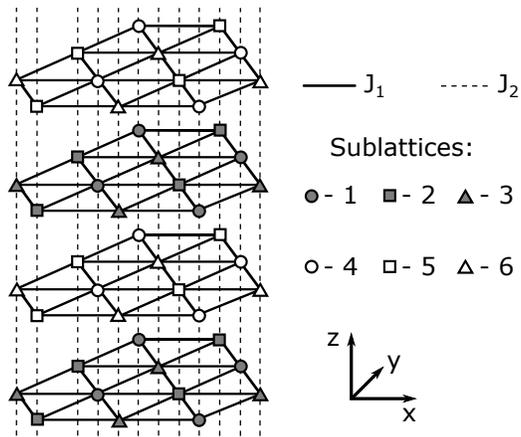}
\caption{ASTIA lattice partitioned into six sublattices marked by different symbols. The solid (dashed) lines represent intralayer (interlayer) interaction $J_1$ ($J_2$).}
\label{fig:ASTIA}
\end{figure}

We consider the ASTIA model described by the Hamiltonian
\begin{equation}
H = - J_1 \sum_{\left\langle i, j \right\rangle}  \sigma_i \sigma_j - J_2 \sum_{\left\langle i, k \right\rangle} \sigma_i \sigma_k - h \sum_i \sigma_i,
\end{equation}
where $\sigma_i=\pm1$ is an Ising spin variable, $J_1<0$ and $J_2<0$ are respectively antiferromagnetic intralayer and interlayer exchange interactions, $h$ is an external magnetic field, and the first and second summations run over the nearest neighbor pairs within and between the layers, respectively. Due to the antiferromagnetic nature of both interactions $J_1$ and $J_2$ it is desirable do decompose the entire lattice into six interpenetrating sublattices, as shown in Fig.~\ref{fig:ASTIA}. The total coordination number is $z=8$ and each spin is coupled to six neighbors from two sublattices ($3+3$) in the same layer and two neighbors from another sublattice in the adjacent layers.

\subsection{Monte Carlo simulations}

In our Monte Carlo (MC) simulations we consider the ASTIA system of the size $V=L_x \times L_y \times L_z=L \times L \times 4L/3$, i.e., $L_z=4L/3$ layers of the size $L \times L$ stacked along the $z$-axis, comprising in total $V=4L^3/3$ spins. For obtaining temperature dependencies of various thermodynamic functions the linear lattice size is fixed to $L=24$ and for the finite-size scaling (FSS) analysis it takes values $L=24,36$, and $48$. In all simulations the periodic boundary conditions are imposed. 

Initial spin states are randomly assigned and the updating follows the Metropolis dynamics. The lattice structure and the short range nature of the interactions enable vectorization of the algorithm. Since the spins on one sublattice interact only with the spins on the other, each sublattice can be updated simultaneously. Thus one sweep through the entire lattice involves just six sublattice updating steps. For thermal averaging, we typically consider $N = 10^5$ MC sweeps in the standard and up to $N = 10^7$ MC sweeps in the histogram MC simulations~\cite{Ferrenberg1,Ferrenberg2}, after discarding another $20$\% of these numbers for thermalization. To assess uncertainty of the calculated quantities, we perform $10$ runs, using different random initial configurations, and the error bars are taken as twice of the standard deviations.

We calculate the enthalpy per spin $e=E/V|J_1|=\langle H \rangle/V|J_1|$, where $\langle \cdots  \rangle$ denotes the thermodynamic mean value, the sublattice magnetizations per spin
\begin{equation}
\label{Magn_i}
m_\alpha = 6 \langle M_\alpha \rangle/V = 6 \Big\langle \sum_{j \in \alpha}\sigma_{j} \Big\rangle/V,\ \alpha=1,2,\hdots,6,
\end{equation}
and the total magnetization per spin
\begin{equation}
\label{Magn_tot}
m = \langle M \rangle/V = \Big\langle \sum_{i=1}^{V}\sigma_{i} \Big\rangle/V.
\end{equation}
The magnetic susceptibility is defined as
\begin{equation}
 \chi_m = \beta (\left\langle M^2 \right\rangle - \left\langle M \right\rangle^2)/V.
\label{eq:SuscM}
\end{equation}
and the specific heat as
\begin{equation}
	C = \beta^2 (\left\langle E^2 \right\rangle - \left\langle E \right\rangle^2)/V,
\label{eq:HeatCapZ}
\end{equation}
where $\beta=1/k_BT$. To measure a degree of the ferrimagnetic ordering within the planes and the antiferromagnetic ordering in the stacking direction, we introduce the order parameters $o_{xy}$ and $o_z$, defined as
\begin{equation}
o_{xy} = \langle O_{xy} \rangle_{z}/L^2 = \langle M_{max}-M_{min}+|M_{med}| \rangle_{z}/L^2,
\label{eq:order_par_oxy}
\end{equation}
and
\begin{equation}
o_z = \langle O_z \rangle/L_z =\Big\langle \sum_{k=1}^{L_z} (-1)^k\sigma_k \Big\rangle_{xy}/L_z,
\label{eq:order_par_oz}
\end{equation}
where $M_{max}$, $M_{min}$, and $M_{med}$ are sublattice magnetizations in each plane with the maximum, minimum, and medium (remaining) values, respectively, and the symbols $\langle \cdots \rangle_{z}$ and $\langle \cdots \rangle_{xy}$ denote the mean values taken over the planes and over the chains, respectively.

To study phase transitions in the present six-sublattice system, we define the order parameter in accordance with Ref.~\cite{Landau} as
\begin{equation}
o = \langle O \rangle/V =\Bigg\langle \frac{\sqrt{3}}{3}\left(\sum_{\alpha=1}^{6} O_\alpha^2 \right)^{1/2}\Bigg\rangle\Bigg/V,
\label{eq:order_par}
\end{equation}
where $O_1 = (M_1 - (M_2+M_3)/2)/2$, $O_2 = (M_2 - (M_1+M_3)/2)/2$, $O_3 = (M_3 - (M_1+M_2)/2)/2$, $O_4 = (O_4 - (M_5+M_6)/2)/2$, $O_5 = (M_5 - (M_4+M_6)/2)/2$, $O_6 = (M_6 - (M_4+M_5)/2)/2$, and the corresponding susceptibility
\begin{equation}
 \chi_o = \beta (\left\langle O^2 \right\rangle - \left\langle O \right\rangle^2)/V.
\label{eq:SuscO}
\end{equation}

In order to calculate the critical exponents and thus determine the order of the transition and also the universality class if the transition is second order, we employ a FSS analysis with the following scaling relations: 
\begin{eqnarray}
	C(L) \propto L^{\alpha/\nu}, \\
	O(L) \propto L^{-\beta/\nu},  \\
	\chi(L) \propto L^{\gamma/\nu}
\label{eq:CrtiExpL}
\end{eqnarray}

\begin{equation}
	\frac{d \left\langle O \right\rangle}{d \beta} = \left\langle O \right\rangle \left\langle E \right\rangle - \left\langle O E \right\rangle \propto L^{(1-\beta)/\nu},
\label{eq:dO}
\end{equation}

\begin{equation}
	\frac{d \ln \left\langle O^2 \right\rangle}{d\beta} = \left\langle E \right\rangle - \frac{\left\langle O^2 E \right\rangle}{\left\langle O^2 \right\rangle} \propto L^{1/\nu},
\label{eq:dlnO}
\end{equation}
where $\alpha,\beta,\gamma$ and $\nu$ are the critical exponents corresponding to the specific heat, the order parameter, its susceptibility and the correlation length, respectively. Having estimated the exponent $\nu$, the inverse critical (N{\'e}el) temperature $\beta_N=1/k_BT_N$ can be obtained from the relation
\begin{equation}
	\beta_{\max}(L) = \beta_N + a_i L^{-1/\nu},
\label{eq:BetaScaling}
\end{equation}
where $\beta_{max}$ is the inverse temperature in the vicinity of the transition point at which various quantities display maxima.

\section{Results and discussion}
\label{results}

\subsection{Ground states}
At zero temperature, the minimum energy states for different fields can be determined directly from the Hamiltonian. In Table~\ref{tab:GS} we present the identified states, showing the sublattice and total magnetizations, as well as the the reduced enthalpy. There are three phases, corresponding to the three field intervals $0 \leq h/|J_1|<-2J_2/|J_1|$ (phase I), $-2J_2/|J_1|<h/|J_1|<6-2J_2/|J_1|$ (phase II) and $6-2J_2/|J_1|<h/|J_1|<\infty$ (fully polarized phase P). 

\begin{table}[t!]
	\centering
	\begin{tabular}{c||c|c|c}
		 $\frac{h}{|J_1|}$ & $( 0, -\frac{2 J_2}{|J_1|} )$ & $( -\frac{2 J_2}{|J_1|}, 6 - \frac{2 J_2}{|J_1|} )$ & $( 6-\frac{2 J_2}{|J_1|}, \infty )$ \\ 
		\hline $\left( \frac{m_1, m_2,  m_3}{m_4, m_5,  m_6}  \right)$ 
		& $\left( \frac{+1,+1,-1}{-1,-1,+1} \right)$ 
		& $\left( \frac{+1,+1,-1}{+1,-1,+1} \right)$ 
		& $\left( \frac{+1,+1,+1}{+1,+1,+1} \right)$ \\
		\hline $m$ & 0 & 1/3 & 1 \\	
		\hline $e$ 
		& $-1 + \frac{J_2}{|J_1|}$
		& $-1 + \frac{J_2}{3|J_1|} - \frac{h}{3|J_1|}$ 
		& $3 - \frac{J_2}{|J_1|} - \frac{h}{|J_1|}$ 
	\end{tabular}
	\caption{Ground states of the ASTIA model in an external field. The respective phases are characterized by the schematic arrangement of the sublattice magnetizations, $m_i$ ($i=1,2,\hdots,6$), the total magnetization, $m$, and the enthalpy, $e$.}
\label{tab:GS}
\end{table}

In the phase I, there is an antiferromagnetic (AF) order within all chains in the $z$-axis direction and the spins in each triangular plaquette (belonging to the neighboring chains) are arranged ferrimagnetically (two parallel and one antiparallel). As one can see, at small fields the enthalpy does not depend on the field values and thus the ground state is expected to be the same as in zero field.

In the phase II, the sublattice magnetization arrangement indicates that spins on one sublattice flip into the field direction and thus two thirds of the chains retain the AF order but remaining one third becomes ferromagnetic (FM). The total magnetization of such a state corresponds to $m=1/3$ and the enthalpy becomes field-dependent.

Finally, in the fully polarized phase, all the spins that are not yet aligned with the field flip to its direction and the total magnetization becomes fully saturated with $m=1$. In the following we set $J_2/|J_1|=-1$ and study the behavior in the field intervals $(0,2),(2,8)$, and $(8,\infty)$.

\subsection{Finite temperatures}

\begin{figure}[t!]
	\centering
	\vspace*{-15mm}
    \subfigure{\includegraphics[scale=0.41,clip]{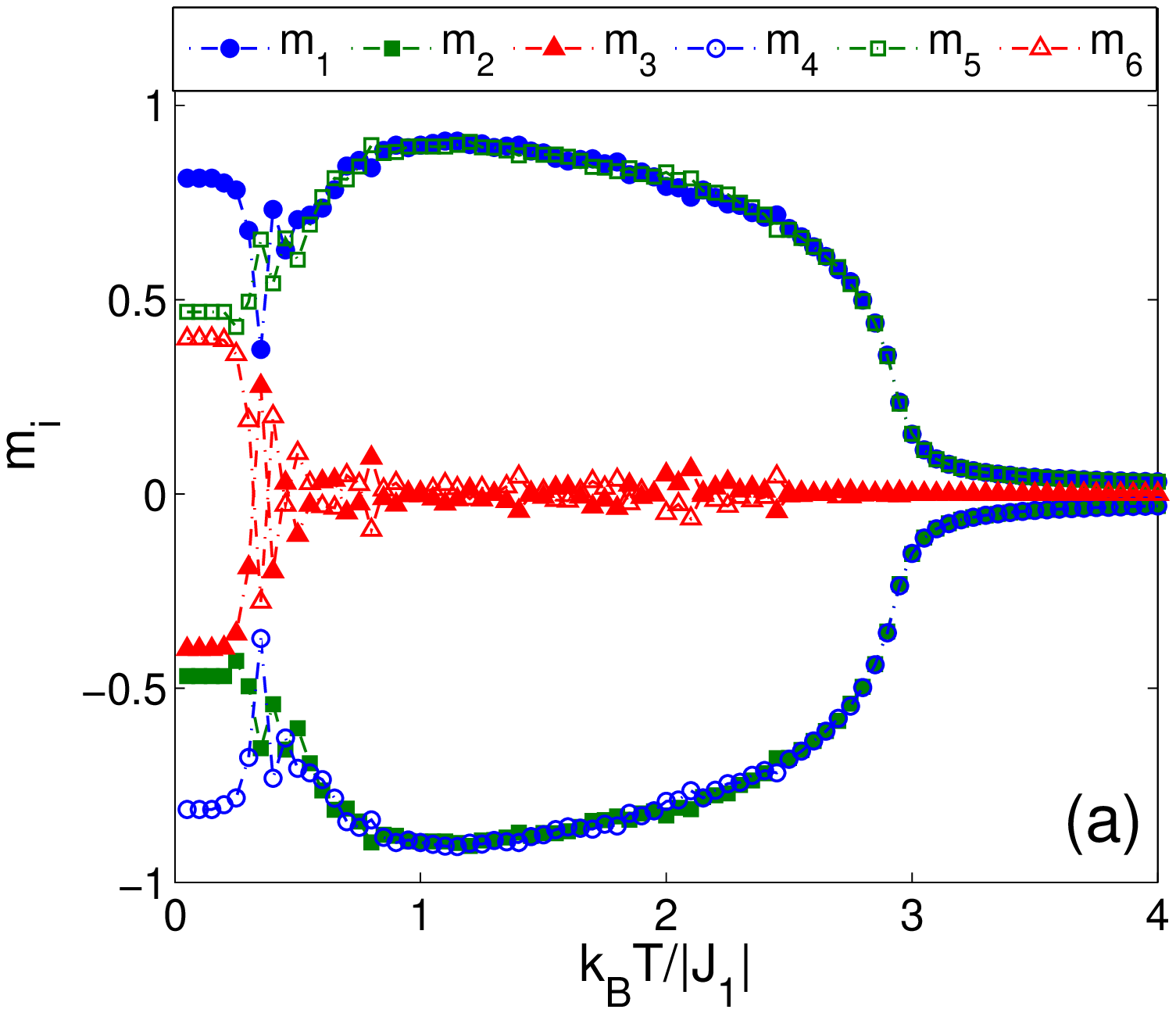}\label{fig:MC_Tmi_L24_H0}}\hspace*{-5mm}
    \subfigure{\includegraphics[scale=0.41,clip]{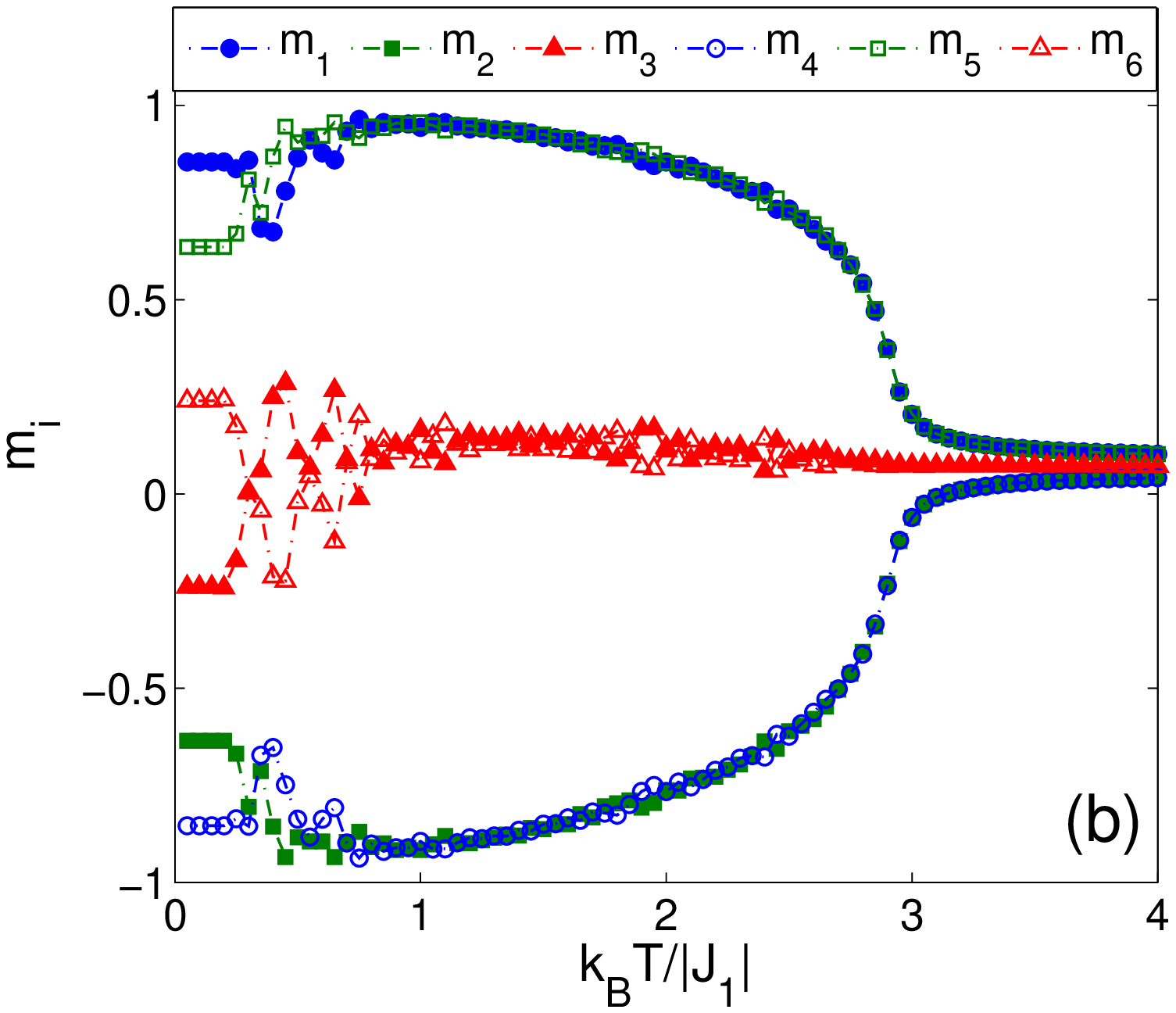}\label{fig:MC_Tmi_L24_H1}}
    \subfigure{\includegraphics[scale=0.41,clip]{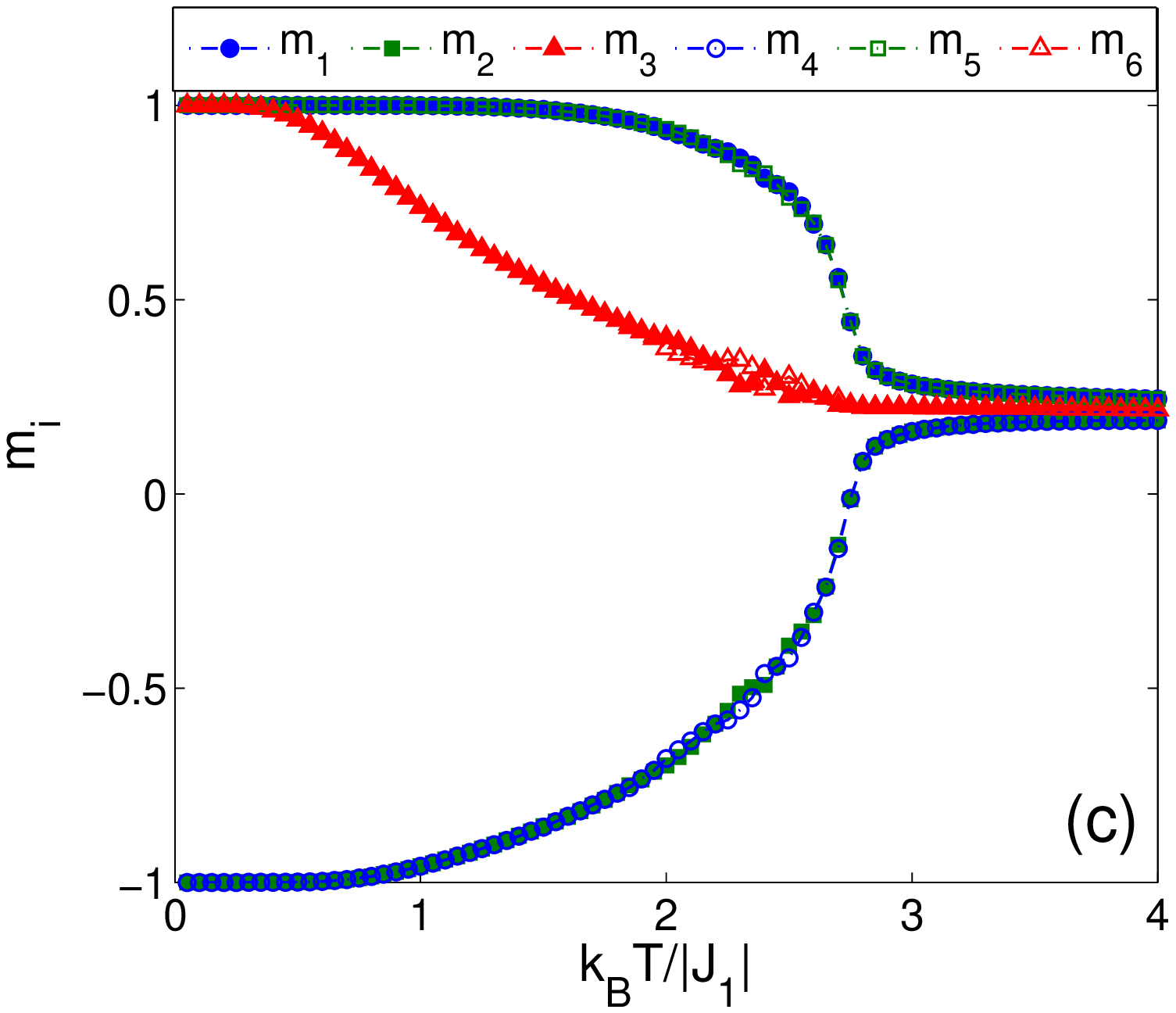}\label{fig:MC_Tmi_L24_H3}}
	\caption{Temperature dependencies of the sublattice magnetizations per spin, $m_i$, $i=1,\hdots,6$, for (a) $h/|J_1|=0$, (b) $h/|J_1|=1$, and (c) $h/|J_1|=3$.}
	\label{fig:MC-T-mi}
\end{figure}

\begin{figure}[t!]
	\centering
    \vspace*{-15mm}
    \subfigure{\includegraphics[scale=0.41]{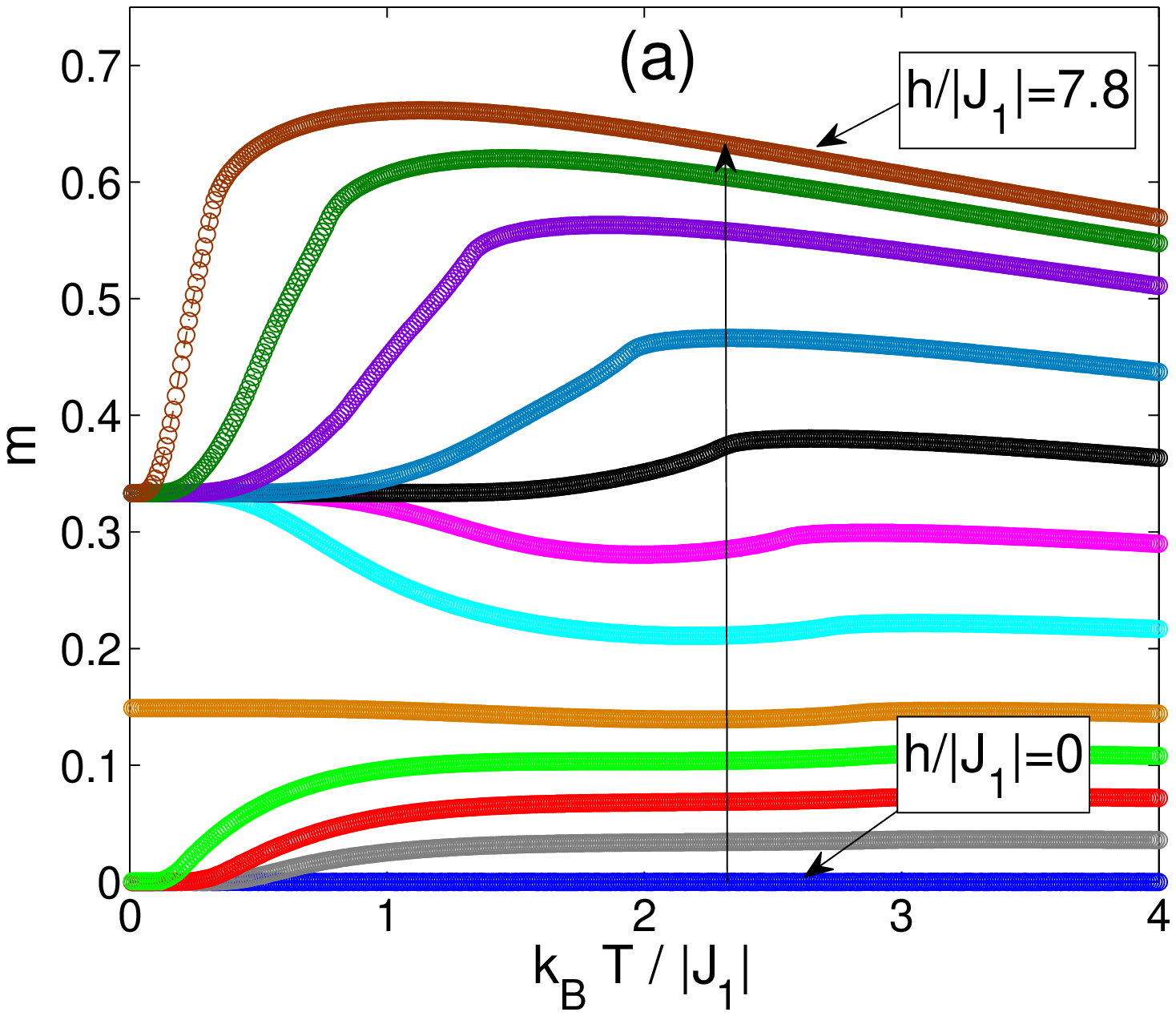}\label{fig:MC_TM}}\hspace*{-5mm}
    \subfigure{\includegraphics[scale=0.41]{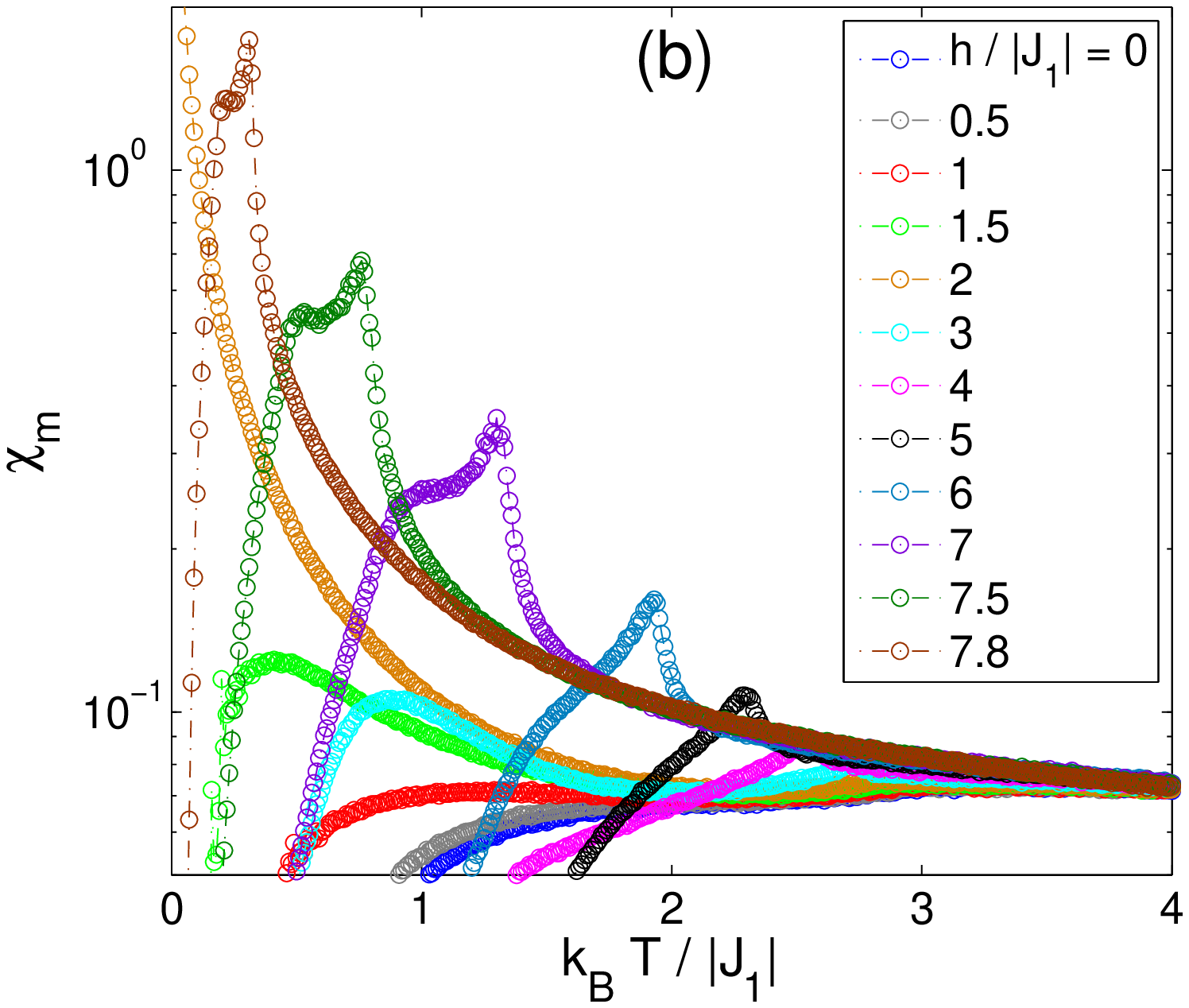}\label{fig:MC_TX}}\vspace*{-5mm}\\
		\subfigure{\includegraphics[scale=0.41]{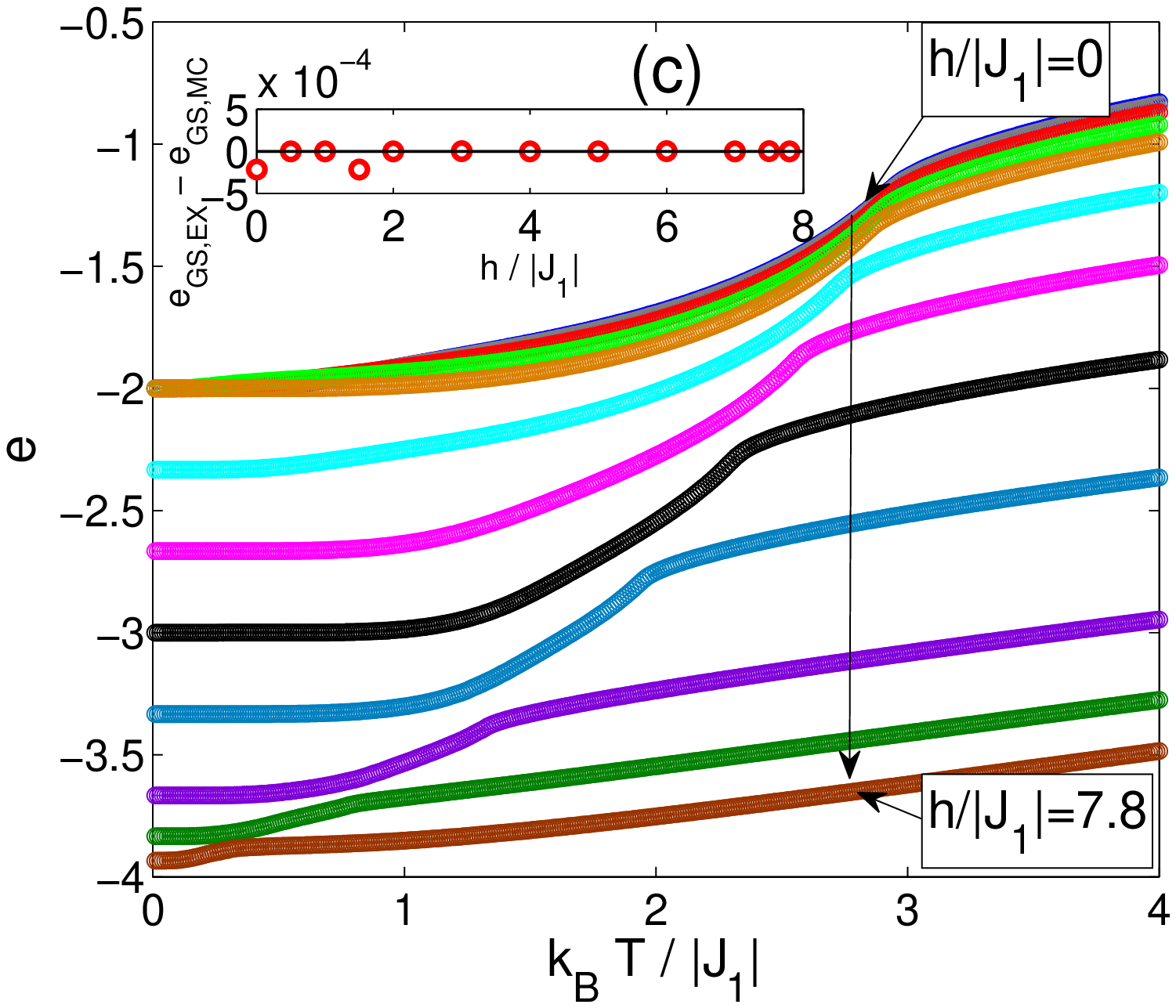}\label{fig:MC_TE}}
\hspace*{-5mm}
    \subfigure{\includegraphics[scale=0.41]{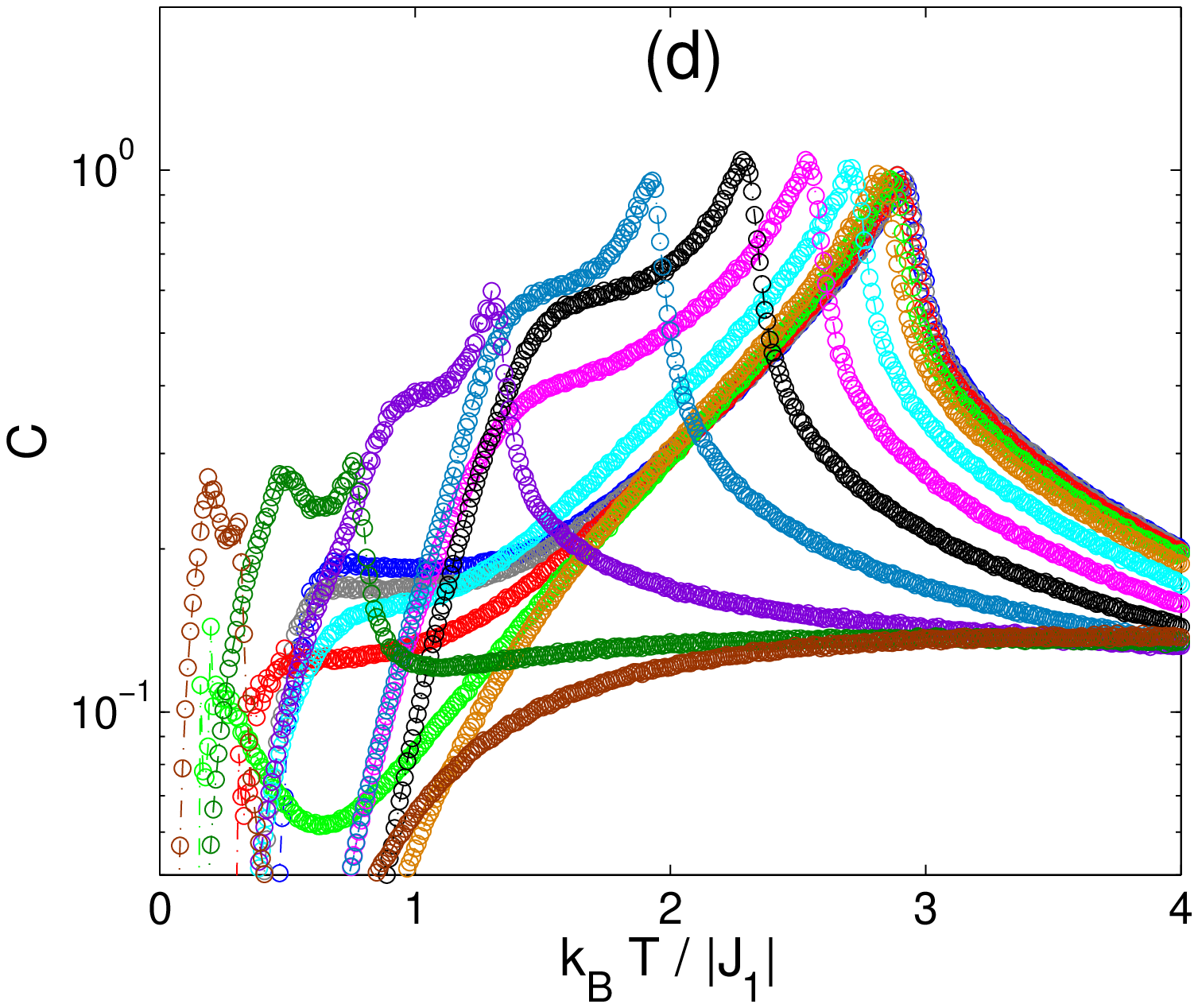}\label{fig:MC_TC}}\vspace*{-5mm}\\
		\subfigure{\includegraphics[scale=0.41]{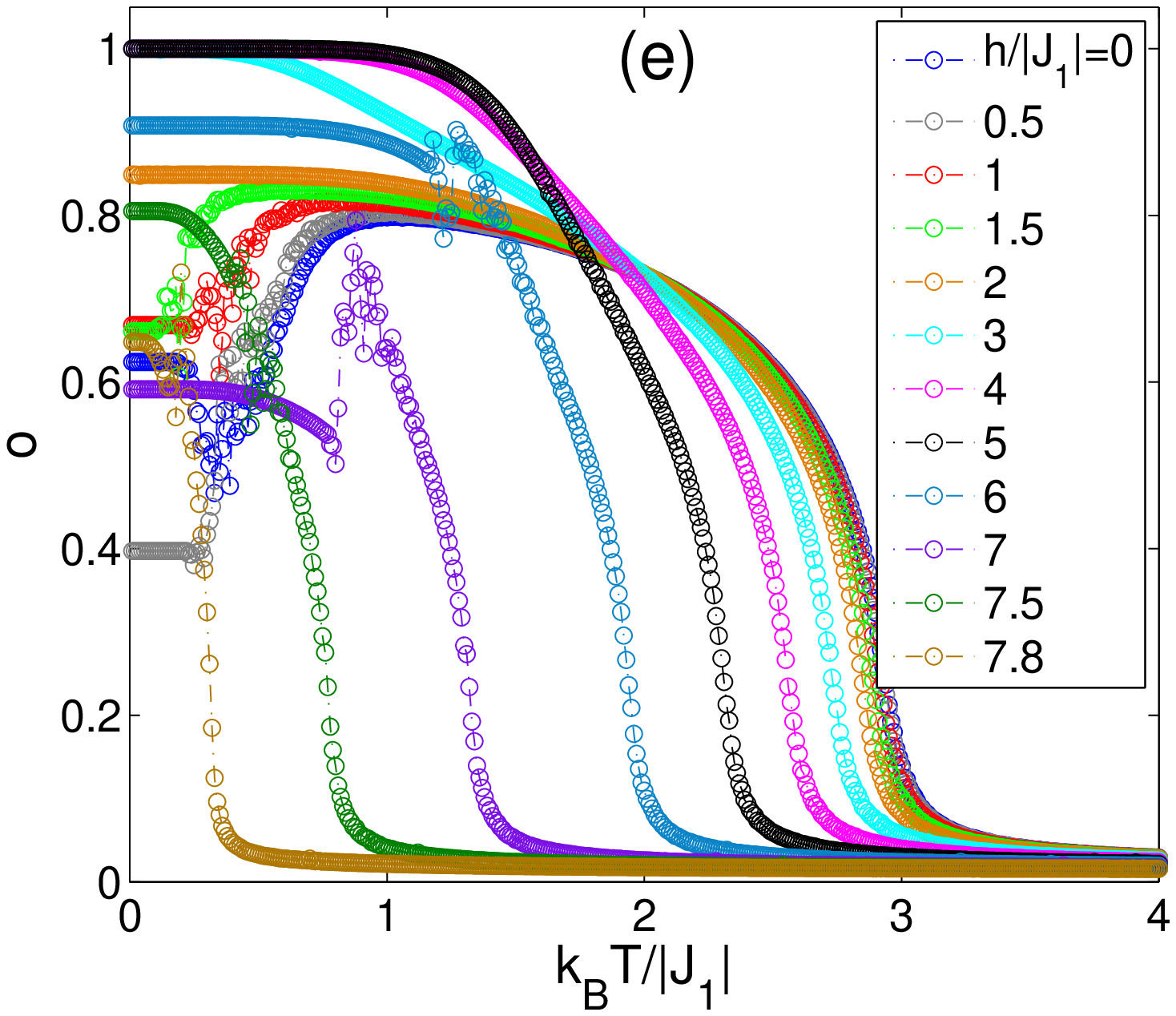}\label{fig:MC_TO}}
	\caption{Temperature dependencies of (a) the total magnetization per spin, (b) the magnetic susceptibility, (c) the enthalpy per spin, (d) the specific heat, and (e) the order parameter $o$, for different fields $h/|J_1|=0,0.5,\hdots,7.8$. The inset in (c) shows the difference between the enthalpy at the lowest simulated temperature and the exact GS value. The arrows in (a) and (c) show respectively increasing and decreasing trends in the magnetization and the enthalpy with the increasing field.}
	\label{fig:MC-T-X}
\end{figure}

In Fig.~\ref{fig:MC-T-mi} we plot the sublattice magnetizations as functions of the temperature, for $h/|J_1|=0,1$ and $3$. In zero field the behavior resembles that of the FSTIA model~\cite{Netz1}, except that there are three more sublattices $m_4,m_5$, and $m_6$ antiferromagnetically coupled to $m_1,m_2$, and $m_3$, respectively, and thus $m_4=-m_1$, $m_5=-m_2$, and $m_6=-m_3$. At the intermediate temperatures one can observe the PD phase with two sublattices in each plane AF ordered and one disordered. In the low-temperature region, all the sublattice magnetizations ``freeze'' without reaching saturation values at zero temperature. The lack of saturation is related to the inherent degeneracy of the phases I and will be discussed in more detail below. Another source of the saturation failure it the kinetic freezing phenomenon, as previously reported in the zero-field FSTIA model~\cite{Netz1,Borovsky}, when a standard single-spin-flip MC simulation is employed. Nevertheless, as we show in the inset of Fig.~\ref{fig:MC_TE}, the lowest-temperature energies reached in our simulations coincide rather well with the true ground-state values. The exceptions are the cases of $h/|J_1|=0$ and $1.5$, where we recorded small deviations of about $(E_{GS,EX}-E_{GS,MC})/N|J_1|=2\times 10^{-4}$, due to the kinetic freezing.

In Fig.~\ref{fig:MC-T-X} we present temperature dependencies of (a) the total magnetization $m$, (b) the magnetic susceptibility $\chi_m$, (c) the enthalpy $e$, (d) the specific heat $C$, and (e) the order parameter $o$, for various values of the field $h/|J_1|$. We can observe the sharp high-temperature and broad low-temperature peaks or shoulders in the response functions. Nevertheless, there are no apparent discontinuities in the magnetization and the energy and the character of the high-temperature peaks of the response functions do not signal any change of the phase transition with the increasing field. The low-temperature anomalies are reflected in the behavior of the order parameter $o$, which due to the degeneracies of the phases I and II failed to reach the saturation value. 

\begin{figure}[t!]
	\centering
\vspace*{-15mm}
    \subfigure{\includegraphics[scale=0.45,keepaspectratio,clip=true]{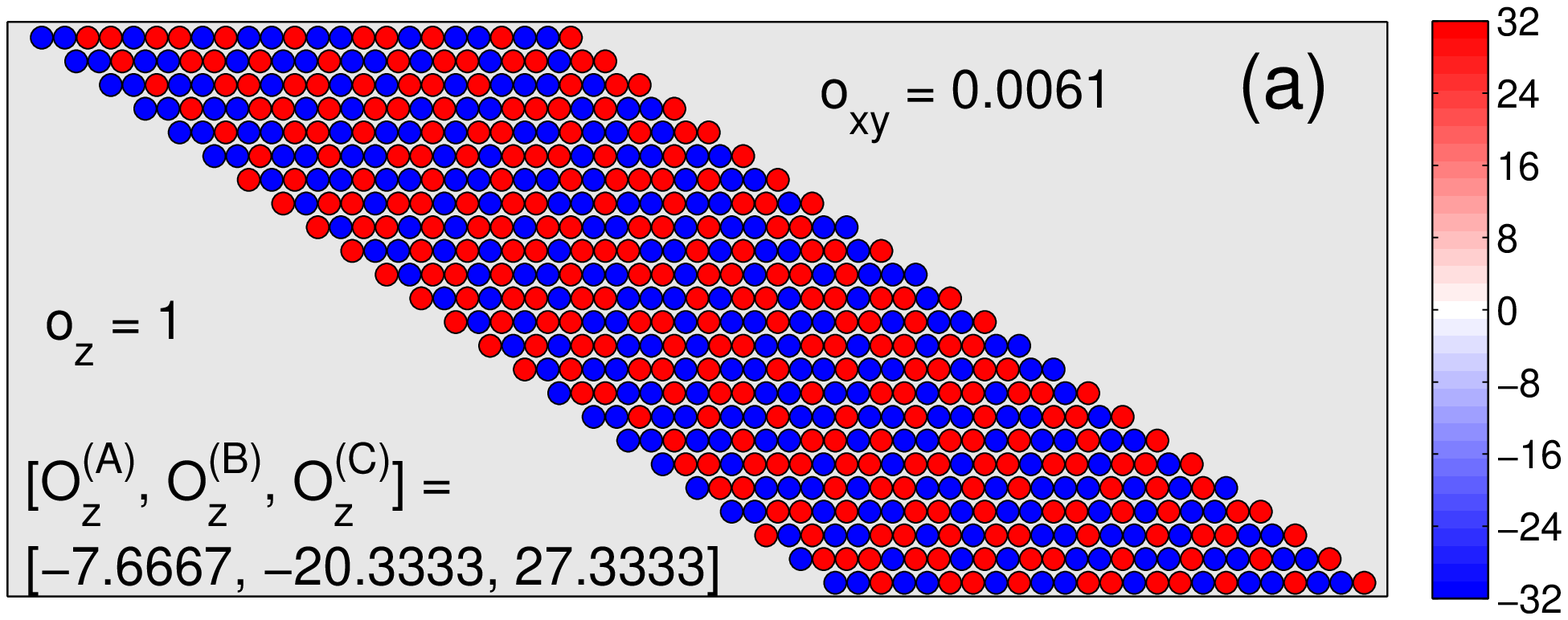}\label{fig:snapshot_h1}}
    \subfigure{\includegraphics[scale=0.45,keepaspectratio,clip=true]{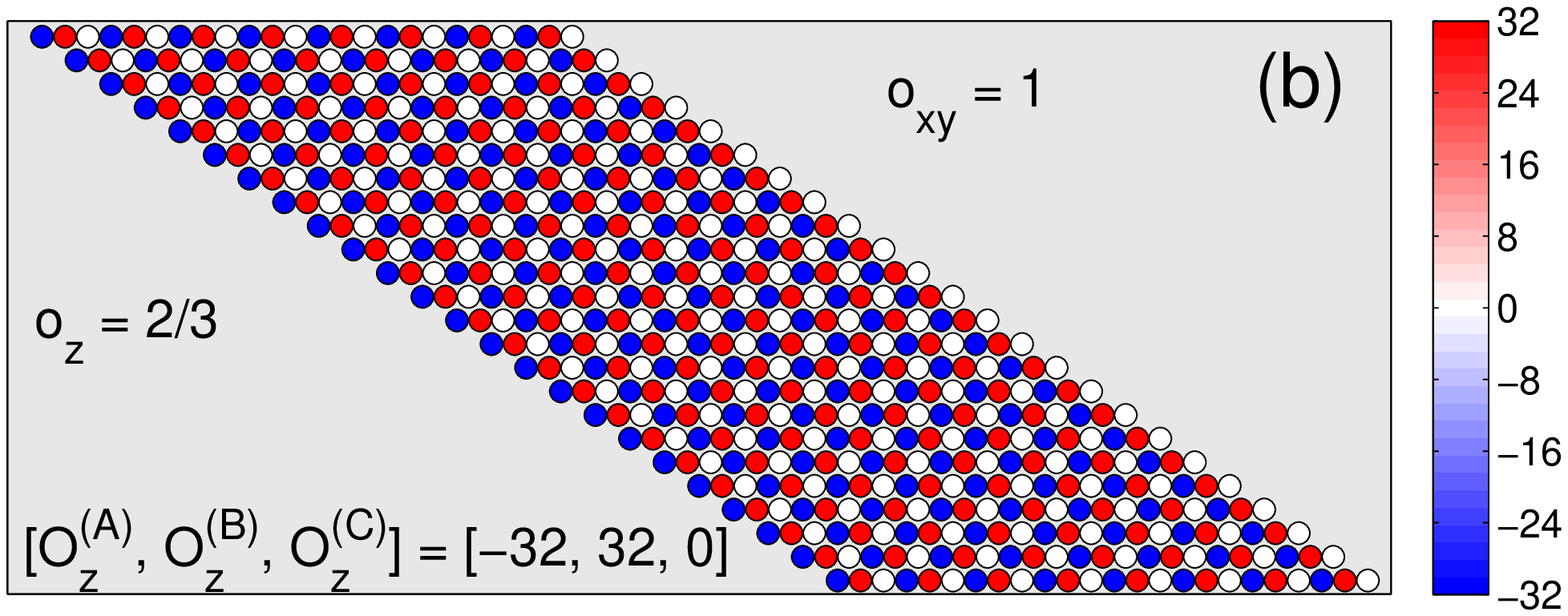}\label{fig:snapshot_h4}}
    \subfigure{\includegraphics[scale=0.45,keepaspectratio,clip=true]{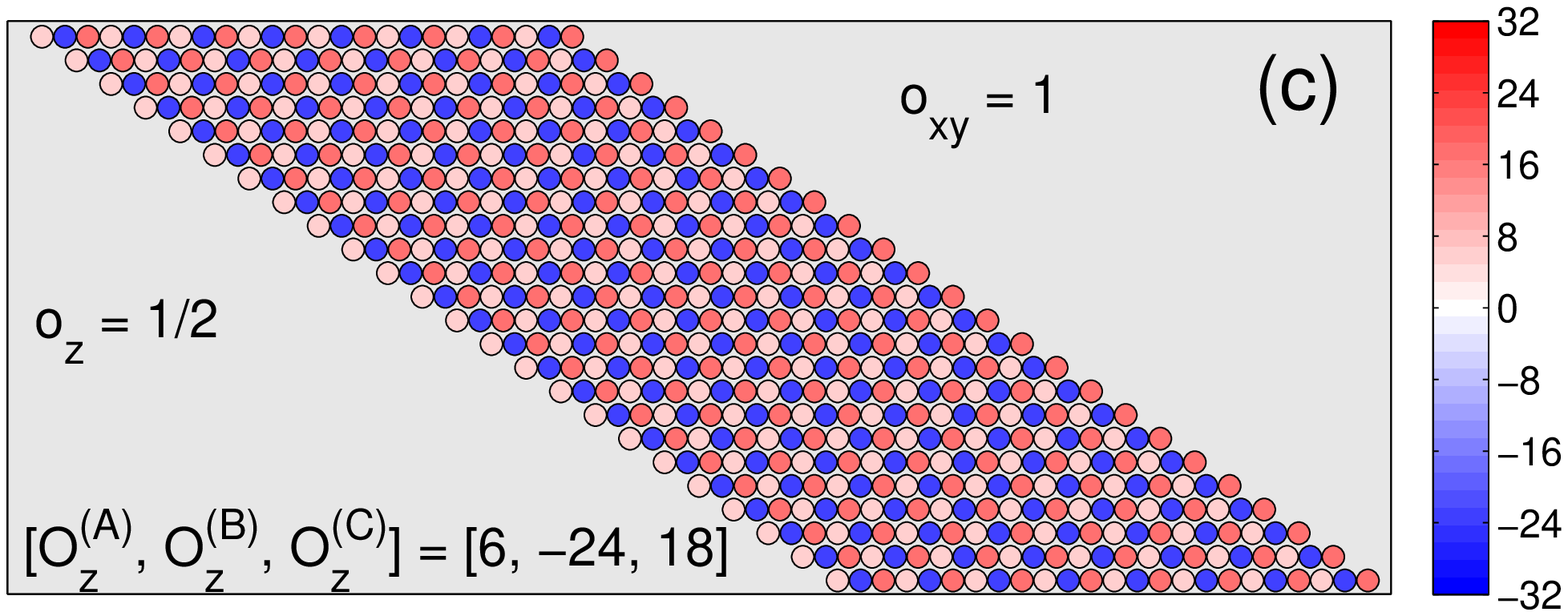}\label{fig:snapshot_h7}}
    \subfigure{\includegraphics[scale=0.45,keepaspectratio,clip=true]{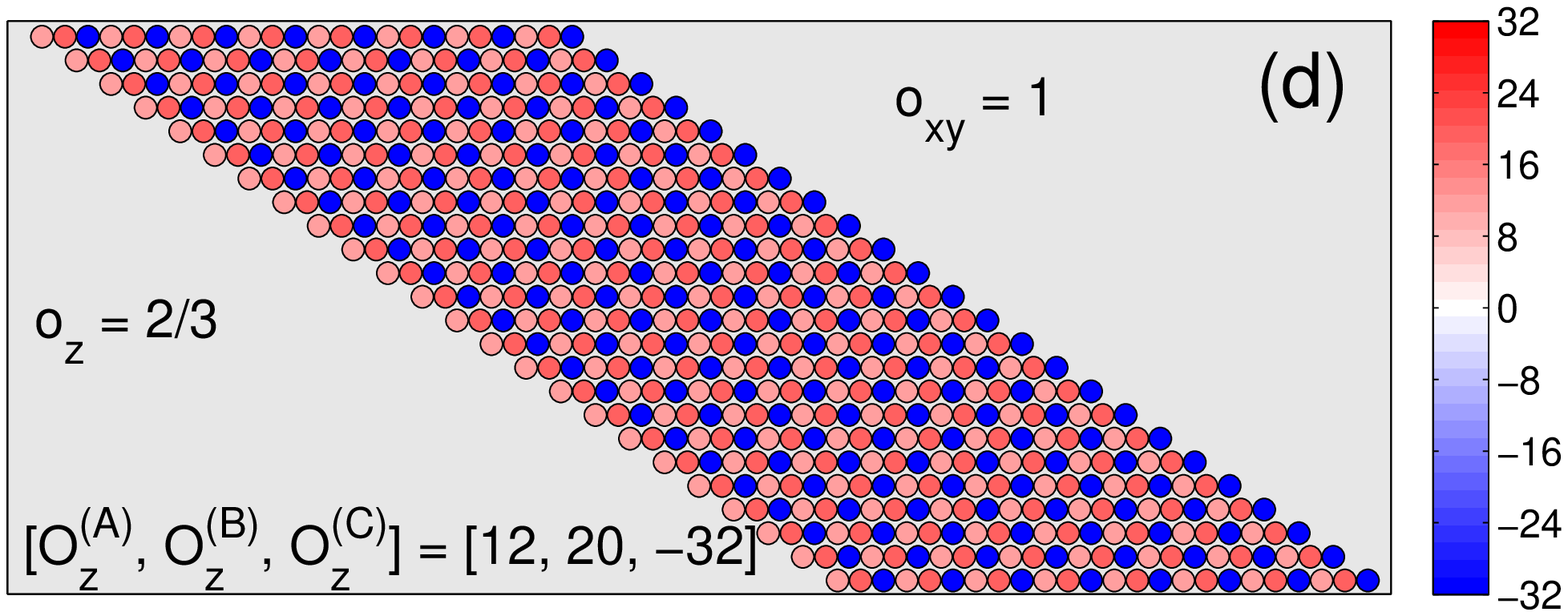}\label{fig:snapshot_h75}}
	\caption{Intraplane and intrachain order parameters $o_{xy}$ and $o_z$, observed close to the ground state ($k_BT_N/|J_1| = 0.01$) for selected field values (a) $h/|J_1|=1$, (b) $h/|J_1|=4$, (c) $h/|J_1|=7$, and (d) $h/|J_1|=7.5$. Individual chains are shown projected onto the $x-y$ plane by circles of different colors representing values of the parameter $o_z$. $O_z^{(A)}$, $O_z^{(B)}$ and $O_z^{(C)}$ give unnormalized values of $o_z$ in the respective sublattices.}
	\label{fig:snapshot}
\end{figure}

Let us study the character of the low-temperature phases I and II in more detail by inspection of MC snapshots taken close to the ground state, at $k_BT/|J_1|=0.01$, where thermal effects are negligible. In Fig.~\ref{fig:snapshot}, we present the snapshots taken at the fields (a) $h/|J_1|=1$, (b) $h/|J_1|=4$, (c) $h/|J_1|=7$, and (d) $h/|J_1|=7.5$, which visualize the ordering within the chains in the stacking direction (the chain order parameter $o_z$) as well as the inplane ordering (the inplane order parameter $o_{xy}$). Individual chains are shown projected to the $x-y$ plane and the degree of their AF ordering is represented by circles of different colors and their intensities: dark (pale) red - full (partial) AF arrangement $(\uparrow\downarrow)$, white - full FM arrangement, and dark (pale) blue - full (partial) AF arrangement $(\downarrow\uparrow)$. The parameters $O_z^{(A)}$, $O_z^{(B)}$, and $O_z^{(C)}$ show the (unnormalized) values of the parameter $o_z$ in the three sublattices of the triangular lattice: A (includes sublattices $1$ and $4$), B (includes sublattices $2$ and $5$), and C (includes sublattices $3$ and $6$).

As one can see, for $h/|J_1|=1$ (phase I) all the chains are perfectly AF ordered ($o_z=1$) but there is no long-range ordering among them ($o_{xy} \approx 0$). The minimum energy condition is satisfied when on each elementary triangular plaquette two chains are parallel and one antiparallel. This state corresponds to the zero-field state and can be considered as a three-dimensional equivalent of the Wannier state, if the fully AF ordered chains are viewed as giant spins. 

On the other hand, the figures in (b), (c) and (d) show examples of rather different spin arrangements in the phase II, all with the intraplane FR LRO. The snapshot in (b) represents the case when all the chains are fully ordered - two thirds of them show AF and one third FM ordering, while the snapshots in (c) and (d) are the examples the FR intraplane LRO without full itrachain ordering. The latter cases apparently result is the unsaturated values of the sublattice magnetizations $m_i$ ($i=1,2,\hdots,6$), as well as the order parameter $o$.

\begin{figure}[t!]
\centering
\vspace*{-15mm}
\includegraphics[scale=0.35,clip]{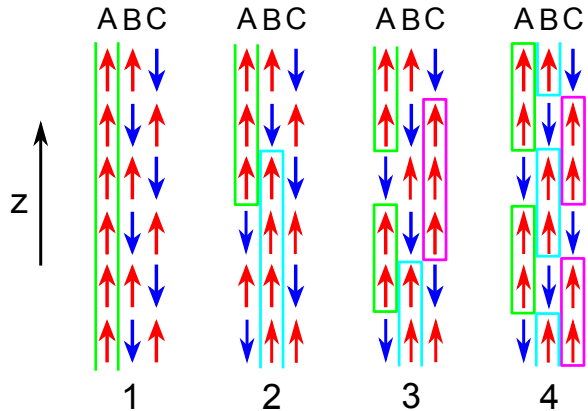}
\caption{Schematic demonstration of the ground-state degeneracy within the phase II. The arrows represent spin orientations in three neighboring chains, with the boxes showing chunks of ferromagnetically arranged chains in the stacking direction.}
\label{fig:ASTIA_degen}
\end{figure}

The mechanism leading to such a behavior is illustrated in Fig.~\ref{fig:ASTIA_degen}. The figure schematically shows spin ordering in the stacking direction in three neighboring chains belonging to the sublattices A, B and C in four degenerate states. The state $1$ corresponds to the snapshot in Fig.~\ref{fig:snapshot_h4}, with the FM chain in the sublattice A marked by the vertical lines. It is easy to verify that if we, for example, swap the lower half of the FM chain with its AF neighbor in the sublattice B (state $2$) the energy remains the same. The FM chain can break into smaller pieces and those can ``migrate'' between different sublattices (states $3$ and $4$) without any energy change. The result is a highly degenerate phase with the lack of saturation of the introduced order parameters.

\begin{figure}[t!]
	\centering
	\vspace*{-15mm}
	  \subfigure{\includegraphics[scale=0.41,clip]{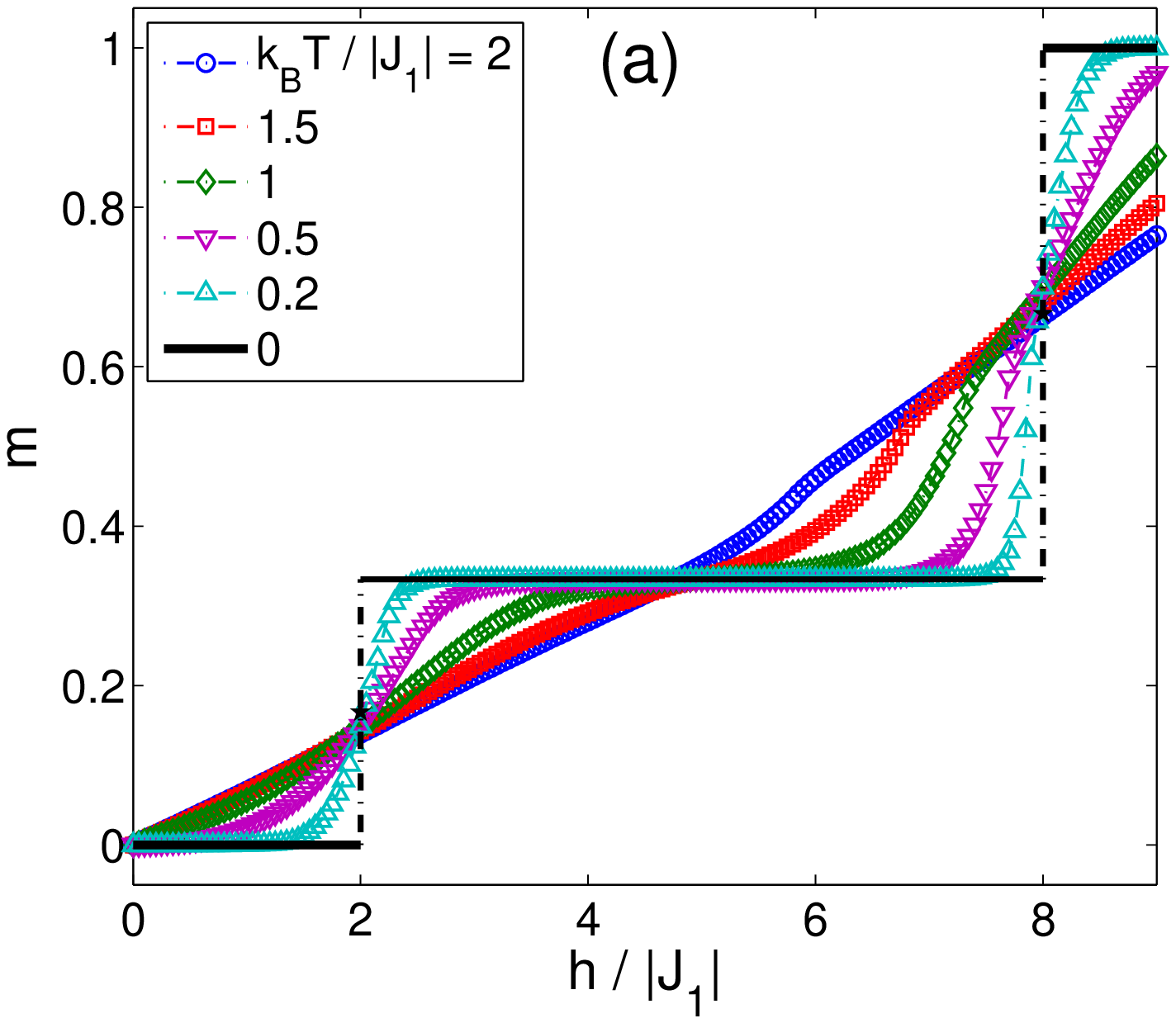}\label{fig:MC_hM}}\hspace*{-5mm}
    \subfigure{\includegraphics[scale=0.41,clip]{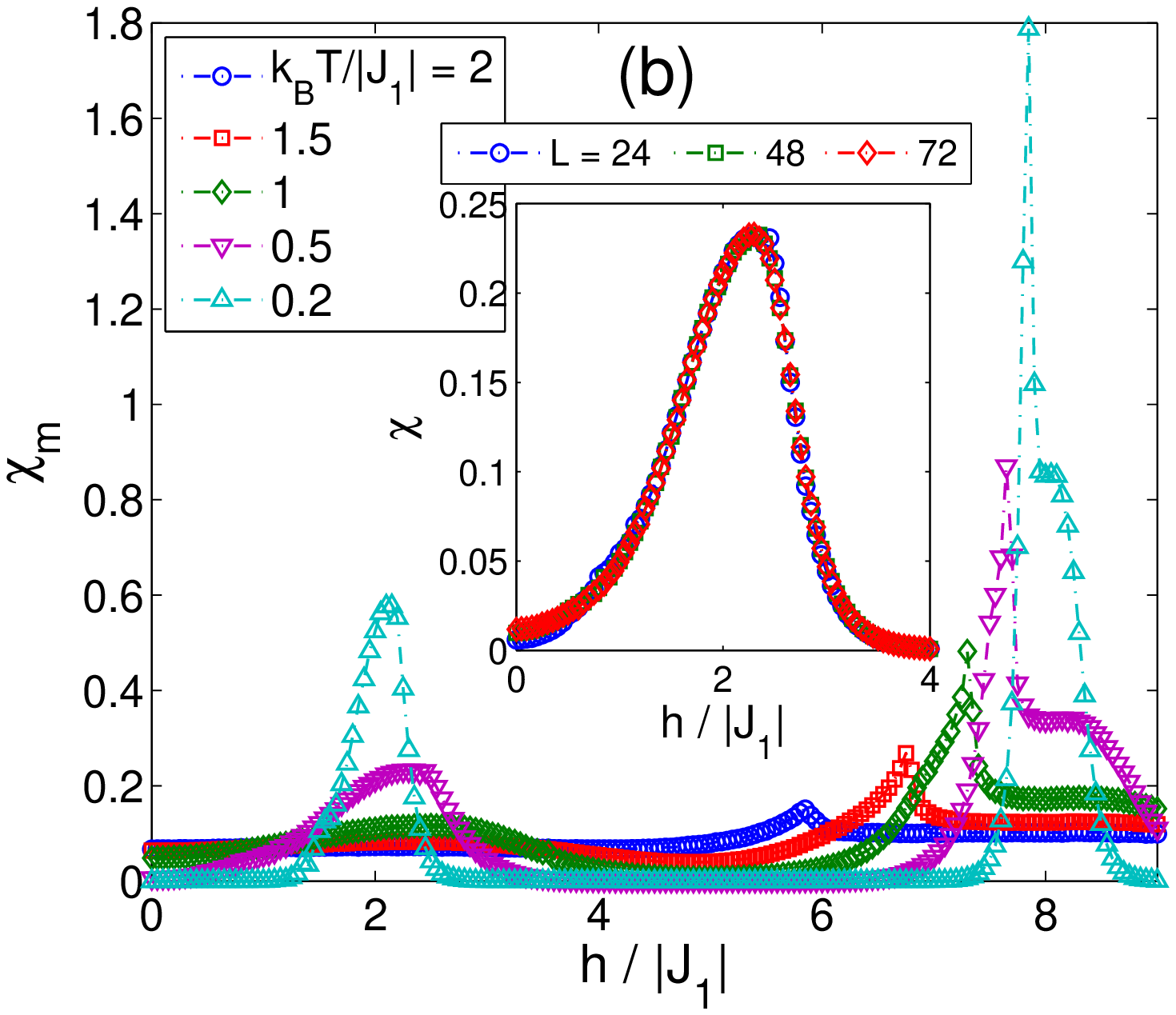}\label{fig:MC_hX}}\vspace*{-5mm}\\
    \subfigure{\includegraphics[scale=0.41,clip]{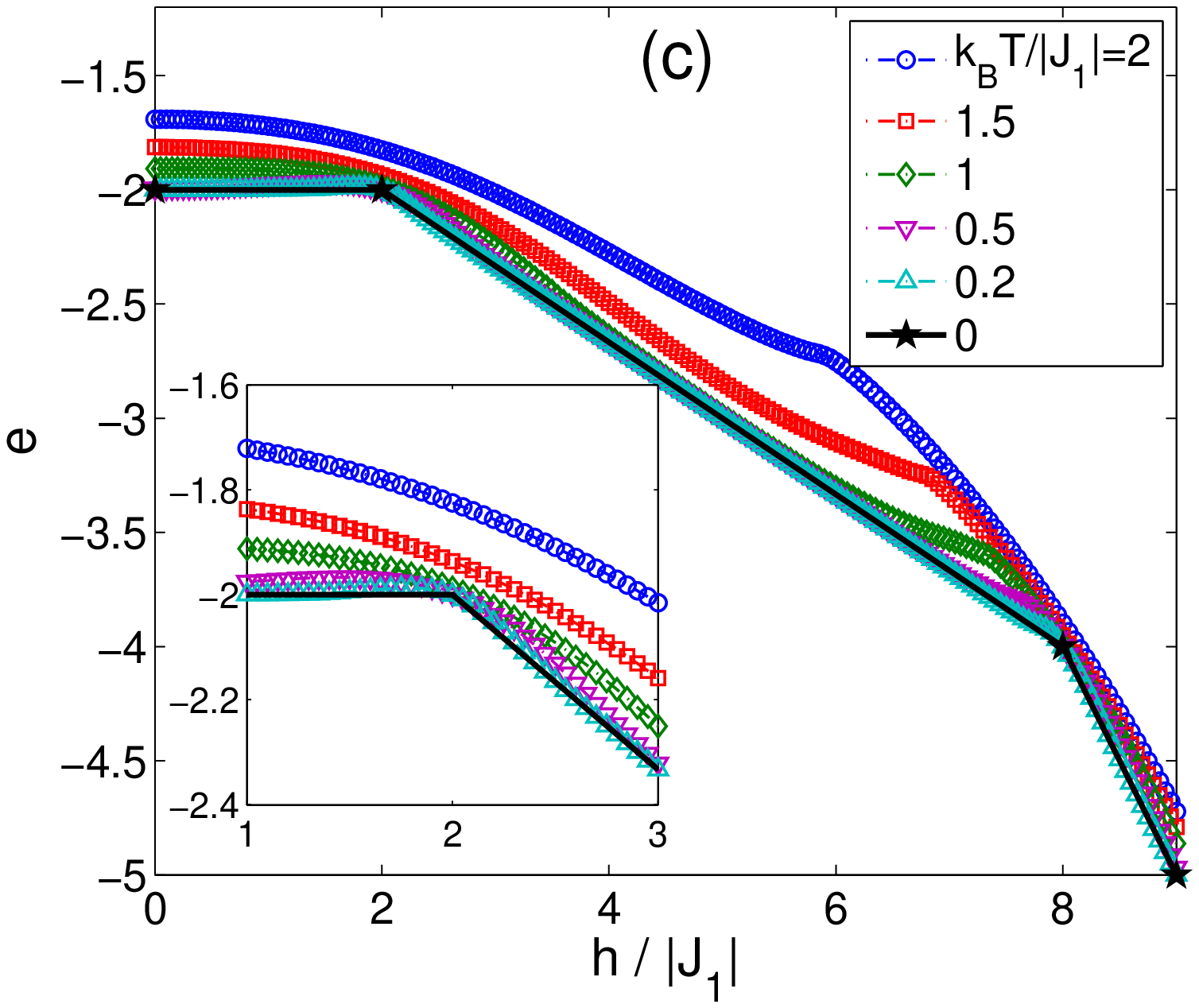}\label{fig:MC_hE}}\hspace*{-5mm}
    \subfigure{\includegraphics[scale=0.41,clip]{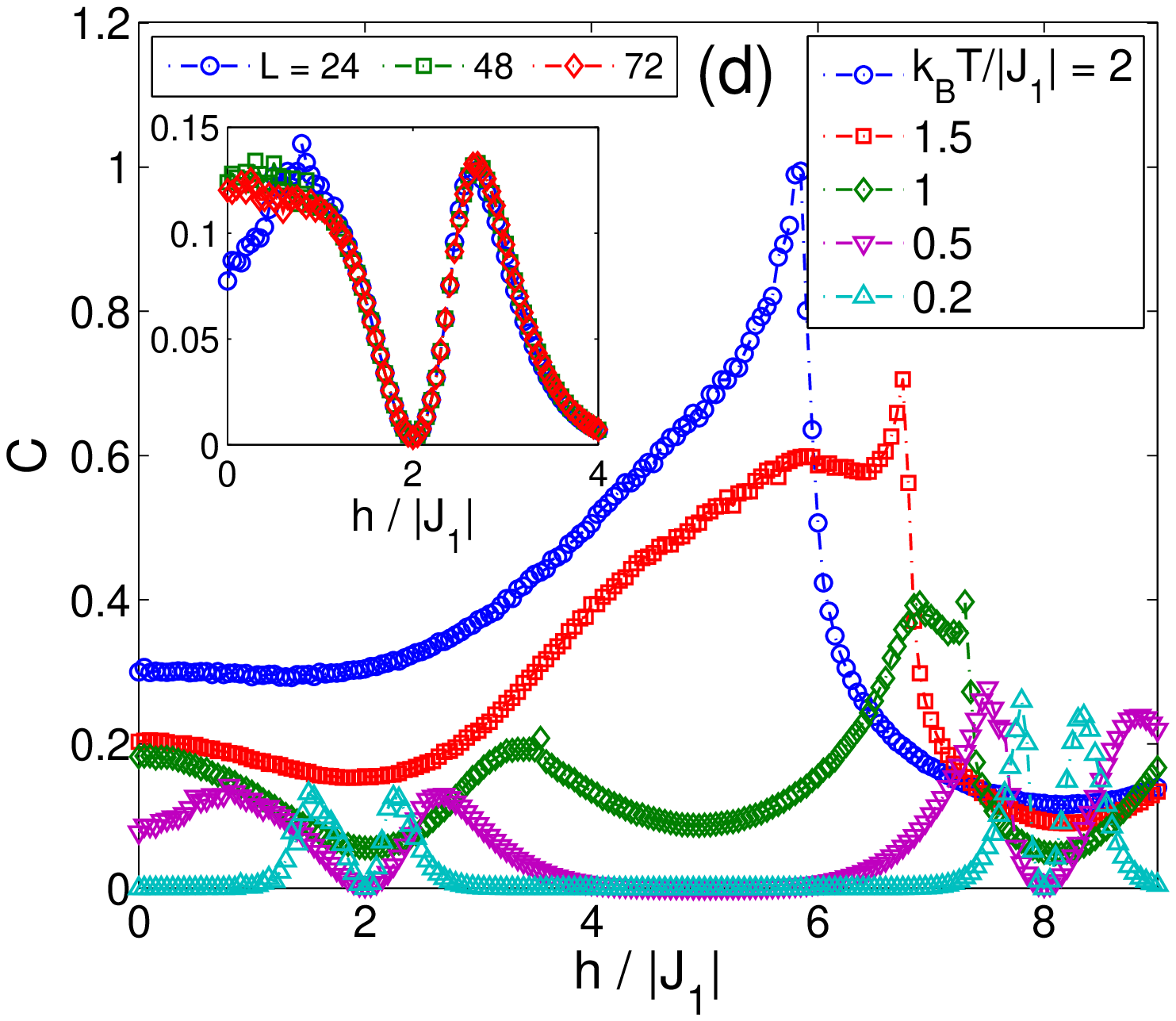}\label{fig:MC_hC}}
	\caption{Field dependencies of (a) the total magnetization per spin, (b) the magnetic susceptibility, (c) the enthalpy, and (d) the specific heat, for different temperatures. The zero-temperature values in (a) and (c) are exact. The insets in (b) and (d) show more detailed views of the quantities for different system sizes.}
	\label{fig:MC_field}
\end{figure}

Fig.~\ref{fig:MC_field} shows the evaluated quantities as functions of the applied field, for different temperatures. In Figs.~\ref{fig:MC_hM} and~\ref{fig:MC_hE} we also include the exact values corresponding to the ground state (bold black lines). Non-differentiability of the enthalpy and discontinuity of the magnetization indicate first-order transitions at $h/|J_1|=2$ and $8$. Nevertheless, at finite temperatures the curves become rounded and the corresponding response functions show a behavior typical for a standard phase transition (sharp peak) only close to $h/|J_1|=8$ but not in the vicinity of $h/|J_1|=2$. Namely, at $h/|J_1|=2$ the low-temperature specific heat becomes suppressed to practically zero and only round-peak anomalies appear from both sides (Fig.~\ref{fig:MC_hC}). The magnetic susceptibility features one round peak close to $h/|J_1|=2$, the height and width of which does not seem to be sensitive the lattice size (see the insets of Figs.~\ref{fig:MC_hX} and \ref{fig:MC_hC}). This behavior makes us believe that the origin of the low-temperature anomalies is not a conventional phase transition but rather linear chain-like excitations. 

\begin{table}[]
	\centering
	\vspace*{-15mm}
	\begin{tabular}{||l||c|c|c|c|c||}
		\hline \hline
		$h/|J_1|$ & $k_B T_N / |J_1|$ & $\alpha$ & $\beta$ & $\gamma$ & $\nu$ \\ \hline \hline
		0 & 2.927 & -0.01(8) & 0.34(2) & 1.34(3) & 0.675(8) \\ \hline
		2 & 2.799 & 0.03(9) & 0.31(3) & 1.33(3) & 0.668(9) \\ \hline
		4 & 2.558 & 0.02(7) & 0.33(2) & 1.33(3) & 0.664(8) \\ \hline
		6 & 1.943 & -0.02(8) & 0.35(2) & 1.32(3) & 0.666(8) \\ \hline
		7 & 1.329 & 0.08(12) & 0.32(3) & 1.29(5) & 0.662(12) \\ \hline
		7.5 & 0.779 & 0.02(21) & 0.28(7) & 1.41(7) & 0.688(18) \\ \hline \hline
		3D XY & - & -0.006 & 0.345 & 1.316 & 0.671 \\ \hline \hline
	\end{tabular}
\caption{Critical exponents and N{\'e}el temperatures for various values of the external field. The last row gives the critical exponents of the three-dimensional XY model universality class.}
\label{tab:FSS}
\end{table}

On the other hand, the high-temperature anomalies in the form of sharp peaks in the response functions appear to indicate standard second-order phase transitions. In order to confirm this presumption based on the standard MC simulation results, we additionally perform the FSS analysis employing the scaling relations~(\ref{eq:CrtiExpL}),~(\ref{eq:dO}) and~(\ref{eq:dlnO}). The estimated N{\'e}el temperatures, obtained from the scaling relation~(\ref{eq:BetaScaling}), and the critical exponents for various fields in the range $0 \leq h/|J_1| \leq 7.5$ are summarized in Table~\ref{tab:FSS}. Due to the well known problem with reliable estimation of the critical exponents $\alpha \approx 0$, their values were obtained indirectly from the Rushbrook relation $\alpha + 2\beta + \gamma =2$. 

\begin{figure}[t!]
\centering
\vspace*{-15mm}
\includegraphics[scale=0.6,clip]{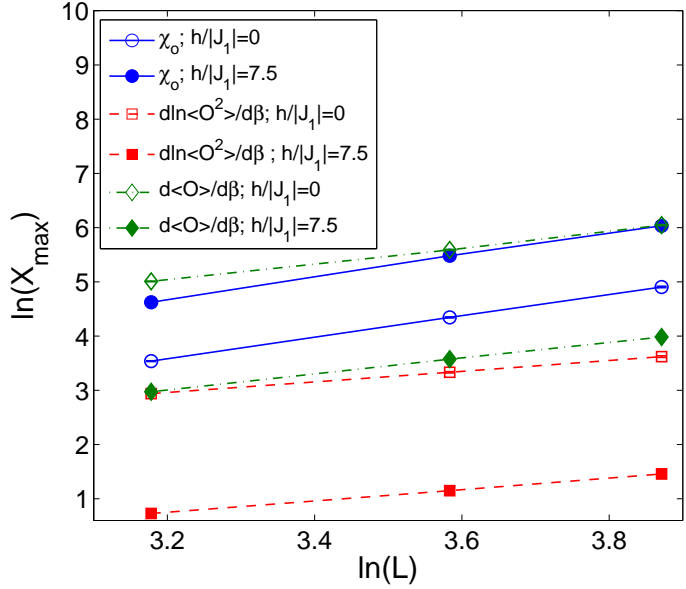}
\caption{FSS of extrema of the quantities $\chi_o$ (blue circles), $d \ln \left\langle O^2 \right\rangle /d\beta$ (red squares) and $d \left\langle O \right\rangle /d\beta$ (green diamonds) at the transition to the paramagnetic state, for $h/|J_1|=0$ (empty symbols) and $7.5$ (filled symbols).}
\label{fig:FSS}
\end{figure}

All the obtained critical exponents are in good correspondence with the three-dimensional XY universality class (see the last row in the table) and, thus, exclude the possibility of the crossover to the first-order transition. This is the case at least for $h/|J_1| \leq 7.5$, however, even the values obtained for $h/|J_1| = 7.5$ are in good accordance with those for $h/|J_1| = 0$ (see also Fig.~\ref{fig:FSS}).  

\begin{figure}[t!]
\centering
\vspace*{-15mm}
\includegraphics[scale=0.6,clip]{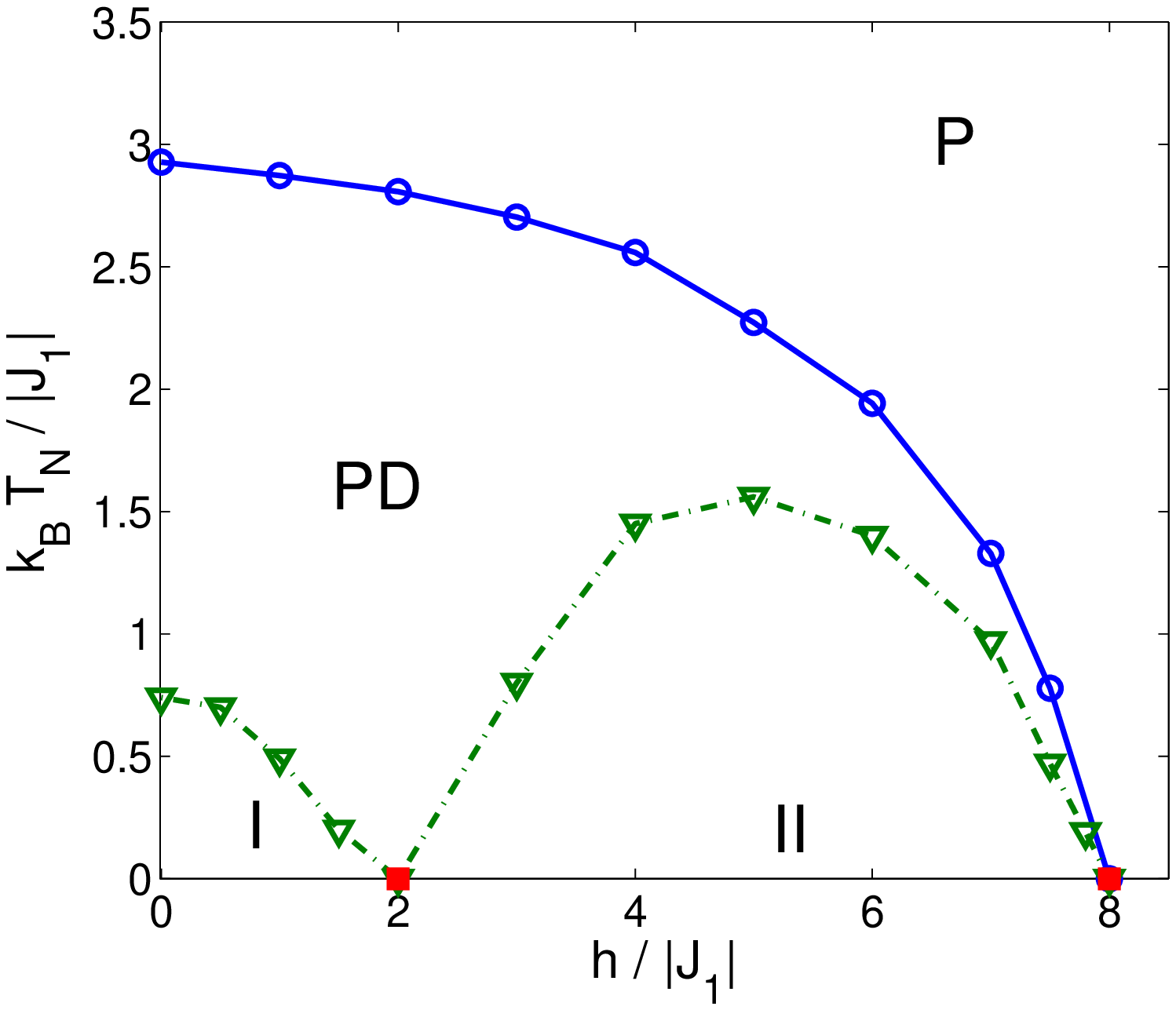}
\caption{Phase diagrams in the $h-T$ parameter plane featuring the paramagnetic (P), the partially disordered (PD) and the highly degenerate I and II phases. The entire P-PD boundary represents second-order phase transitions. The downward triangles mark the low-temperature anomalies observed in the specific heat at the PD-I and PD-II crossovers. The red solid squares show the ground-state transition points between the phases I and II at $h/|J_1|=2$ and from II to the fully polarized P phase at $h/|J_1|=8$.}
\label{fig:PD_MC}
\end{figure}

Finally, in Fig.~\ref{fig:PD_MC} we present the phase diagram. The order-disorder phase boundary (circles) represents second-order phase transitions belonging to the 3D XY universality class. The low-temperature boundaries (downward triangles) are determined from anomalies in the specific heat and represent crossovers to the highly degenerate phases I and II.

\section{Summary and conclusions}
\label{summary}

We studied the stacked triangular Ising antiferromagnet (ASTIA model) with antiferromagnetic (AF) interactions both within the triangular planes ($J_1<0$) as well as in the stacking direction ($J_2<0$) by Monte Carlo (MC) simulations. 

At zero temperature we identified three ground-state phases, corresponding to the three field intervals $0 \leq h/|J_1|<-2J_2/|J_1|$ (phase I), $-2J_2/|J_1|<h/|J_1|<6-2J_2/|J_1|$ (phase II) and $6-2J_2/|J_1|<h/|J_1|<\infty$ (phase P). The phase I is characterized by a full AF spin arrangement in the stacking direction but no long-range ordering (LRO) within the planes (Wannier-like phase). On the other hand, in the phase II the system shows a ferrimagnetic $(\uparrow\uparrow\downarrow)$ LRO within the planes but no LRO in the stacking direction. Thus, both the phases I and II are highly degenerate but in different directions. The phase P represents a fully polarized phase with all spins pointing in the field direction. 

At finite temperatures we limited our considerations to the case of $J_2/|J_1|=-1$. The results indicated the presence of only one phase transition within the entire interval $0 \leq h/|J_1|< 8$, which is of second order belonging to the 3D XY universality class. We note, that this behavior is quite different from the FSTIA model, which shows a crossover to the first-order regime at finite fields. There are also anomalies in the response functions at lower temperatures but their character (broad bumps and shoulders insensitive to finite-size effects) do not indicate the occurrence of any conventional phase transition or even appearance of a new intermediate phase, as suggested by some previous approximate approaches~\cite{Plumer0,Netz2}, but rather linear-chain-like excitations.

In the current study we focused on the isotropic interaction case $J_1=J_2$. In order to better understand unusual properties of the quasi one-dimensional Ising-like antiferromagnets $\rm{CsCoCl}_3$ and $\rm{CsCoBr}_3$ in the future work it would be interesting to extend the present investigation to the dimensional crossover region of $|J_2|>>|J_1|$. The dimensional crossover phenomena in such a frustrated spin system would also be of theoretical interest.

\section*{Acknowledgments}
This work was supported by the Scientific Grant Agency of Ministry of Education of Slovak Republic (Grant No. 1/0331/15) and the scientific grants of the Slovak Research
and Development Agency provided under contract No.~APVV-16-0186 and No.~APVV-14-0073.

\end{document}